# Synchronization in a terahertz ring network

Elman Shahverdiev
Azerbaijan National Academy of Sciences Institute of Physics

H. Javid Avenue, 31
Baku- 1143, Azerbaijan

E-mail: e.shahverdiev@physics.science.az

## Abstract

The simplest case of a ring topology is numerically investigated using the Terahertz modeling. Numerical simulations demonstrate high level degree of complete synchronization. Some security implications for the Terahertz communication and computer networks are emphasized.

## 1 Introduction

Chaos control, including chaos synchronization is of pivotal importance in science, technology and nature, see, e.g. [1-4] and references there-in.

Synchronization between systems can help to achieve higher power lasers. This is especially true for the synchronization between Terahertz sources, emitting frequencies in the diapason 0.3-3 Terahertz [2-4]. Terahertz sources sometimes called Terahertz lasers. In recent years it was established that synchronization between Terahertz sources could be potentially helpful to achieve milli-Watt powers. Such powers can be vital in creating adequate powers for practical applications. Synchronization between thousands and thousands Josephson junctions present in high temperature superconductors such as BSCCO (Bismuth Strontium Calcium Copper) could be helpful in achieving this goal [2-5]. Additionally, such synchronization is of immense use to create mobile, small size, cost effective Terahertz devices, which can be used in remote security screening, in detecting fake painting, plastic land mines, non-destructive cancer diagnostics etc., see, e.g. [2-5].

Many different synchronization states have been studied, see, e.g. [1]. Complete or identical synchronization [6-7] was the first to be discovered and is the simplest form of synchronization in chaotic systems. Among others, some types of synchronization include: phase synchronization [8], lag synchronization [9], inverse synchronization [10] (some researchers also use the term anti-phase [11] synchronization), generalized synchronization [12], cascaded and adaptive synchronization [13], dual and dual-cross synchronization [14], anticipating synchronization [15-16], etc.

Synchronization in complex systems is of a certain importance in governing and performance improving point of view [17-18].

As synchronization is associated with communication, the study of existence and stability conditions for synchronization is of paramount importance in networks. Synchronization is important in chaos-based communication system to decode the transmitted message [1-2].

While focusing on the positive side of the chaos synchronization, one should not forget about the situations when synchronization between interacting systems could be quite harmful. For example, in epileptic patients synchronization between neurons could be the reason for epilepsy seizures [15]. Anticipating synchronization could be quite helpful for diagnostic purposes by anticipating epileptic seizures [15-16].

Due to the finite speed of information propagation between the interacting systems, feedbacks and memory effects, etc time delay systems are widespread in the scientific and technological areas, in the natural world [1]. Also such systems can be used to model space-time processes, see e.g. [19]. Because time delay systems are in fact infinite dimensional (as initial conditions are given on the interval and number of points is infinite) these systems can be used to describe partial differential equations. Time delay systems are also capable to generate hyperchaos - property very attractive from the application viewpoint in chaos-based communication, see, e.g. [20-21] and references there-in.

This paper studies chaos synchronization in one of the simplest cases of the network(s) based on the Terahertz system- one of the famous interdisciplinary model of chaotic dynamics in time delay systems [2-4].

The importance of this paper is in the spanning a bridge between chaos synchronization, computer and Terahertz laser network(s). An extra layer of security could be provided by such an approach to communicating data packets between the computers and between Terahertz lasers.

In this paper we consider a ring topology [22] connection between three Terahertz systems. Topology of the connection between systems is one of the simplest. We numerically investigate the possibility of complete synchronization with the highest degree of correlation between the Terahertz systems linked in a ring topology. These findings are of certain importance in communication between computers, in obtaining high power lasers, in communication between Terahertz lasers (sources).

## 2 Ring topology of terahertz sources and synchronization

In this paper nodes are described by the Terahertz sources coupled with time delay. Lines connecting the nodes are called links. Fig.1 shows the schematic description of the simplest case of the ring topology [22].

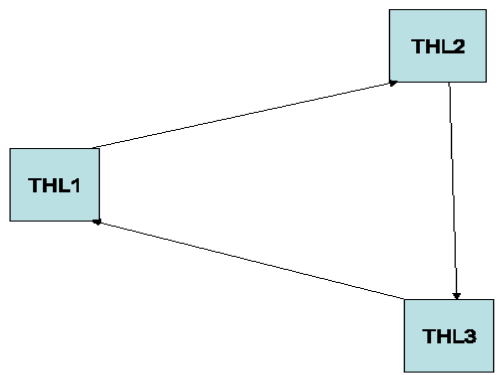

**Fig. 1** Schematic description of the simplest case of the ring topology for the Terahertz lasers. THL 1, 2 and 3 stand for Terahertz Lasers 1, 2 and 3, respectively.

Consider the following set of Terahertz sources forming the simplest ring topology:

$$\frac{d\phi_1}{dt} = \Psi_1, \tag{1}$$

$$\frac{d\Psi_1}{dt} = -\beta\Psi_1 - \sin\phi_1 + i_{dc} + i_0 \cos\varphi_1 - \alpha(\Psi_1 - \Psi_3(t-\tau)), \tag{2}$$

$$\frac{d\varphi_1}{dt} = \Omega, \tag{3}$$

$$\frac{d\phi_2}{dt} = \Psi_2, \tag{4}$$

$$\frac{d\Psi_2}{dt} = -\beta\Psi_2 - \sin\phi_2 + i_{dc} + i_0 \cos\varphi_2 - \alpha(\Psi_2 - \Psi_1(t-\tau)), \tag{5}$$

$$\frac{d\varphi_2}{dt} = \Omega, \tag{6}$$

$$\frac{d\phi_3}{dt} = \Psi_3, \tag{7}$$

$$\frac{d\Psi_3}{dt} = -\beta\Psi_3 - \sin\phi_3 + i_{dc} + i_0 \cos\varphi_3 - \alpha(\Psi_3 - \Psi_2(t-\tau)), \tag{8}$$

$$\frac{d\varphi_3}{dt} = \Omega, \tag{9}$$

Where $\phi_1, \phi_2, \phi_3$ the phase differences of the superconducting order parameter are across the junctions 1, 2, 3, respectively; $\Psi_1, \Psi_2, \Psi_3$ describe dynamics of $\phi_1, \phi_2, \phi_3$ respectively;

$\beta$ is called the damping parameter and is related to McCumber parameter $\beta_c$: $\beta^2 \beta_c = 1$; $i_{dc}$ is the driving the junctions direct current; ; $i_0 \cos(\Omega t + \theta)$ is the driving ac (or rf) current with amplitudes $i_0$, frequencies $\Omega$ and phases $\theta$; $\tau$ is the coupling time delay between Josephson junctions 1-2, 2-3, 3-1; $\alpha$ is the coupling strength between junctions. Note that in (1-9) direct current and $ac$ current amplitudes $i_0$ are normalized with respect to the critical currents for the relative junctions; $ac$ current frequencies $\Omega$ are normalized with respect to the Josephson plasma frequency. Dimensionless time is normalized to the inverse plasma frequency.

It is worth mentioning an elegant derivation of Eqs.(1-3) in [23]. Referring to [23] we also give some typical values for the Josephson junctions. A typical value of emitted frequency by the Josephson junction is of the order of $10^{11} Hz$; the value of the McCumber parameter $\beta_c$ ranges from $10^{-6}$ to $10^6$; typical values of $\beta$ is in the diapason $10^{-3}$-$10^3$; for high temperature superconductors BSCC0 $\beta << 1$ [2]. The coupling strength between the junctions changes in the range $10^{-4}$ -$10^2$; The Josephson junction plasma frequency is on the order of 100GHz; Dimensionless $i_d$ and $i_0$ changes from $10^{-6}$ to $10^1$; non-dimensional $\Omega$ is around $10^{-2}$ and $6 \times 10^{-1}$. One also should emphasize that the typical value for the length scale of the Josephson junctions is around $1 \mu m$.

We underline that equations describing Josephson junctions dynamics are of multidisciplinary nature, see, e.g. [2-4] and references there-in.

We also emphasize that the coupling topology considered in this paper are completely different from the topology investigated in [2-4], where Terahertz lasers were governed by the central Terahertz source. In this paper we investigate the simplest case of the ring topology formed by the Terahertz lasers.

## 3 Numerical simulations and discussions

This section numerically demonstrates the principal possibility of complete chaos synchronization between Terahertz systems. Numerical simulations are conducted with the help of MATLAB software (R2008b). Synchronization quality is characterized by the cross-correlation coefficient C between the dynamical variables.

We study chaos synchronization between Eqs. (1-9) using the correlation coefficient. $C = 1$ means perfect complete synchronization. In simulating Eqs (1-9) we take the parameters of equations equal to show the principal possibility of complete synchronization between Terahertz Lasers. As it is proven in the scientific literature best quality synchronization occurs if the system parameters are equal. We will also guided by this conjecture. But as in the real-world parameters can differ, we simulated Eqs. (1-9) with parameter mismatches 3-5%. Still, we have obtained close to 100 % of correlation between the dynamics of Terahertz Lasers.

We simulate Eqs. (1-9) for the following parameters:

$\beta = 0.25, \alpha = 0.45, \tau = 0.35, i_{dc} = 0.3, i_0 = 0.7, \Omega = 0.6, \theta = 0$.

with initial states
$(\psi_1,\phi_1,\varphi_1)=(1.01,2.05,0),(\psi_2,\phi_2,\varphi_2)=(1.02,2.01,0),(\psi_3,\phi_3,\varphi_3)=(1.05,1.99,0),$
Fig. 2 depicts dynamics of variable $\Psi_2$.

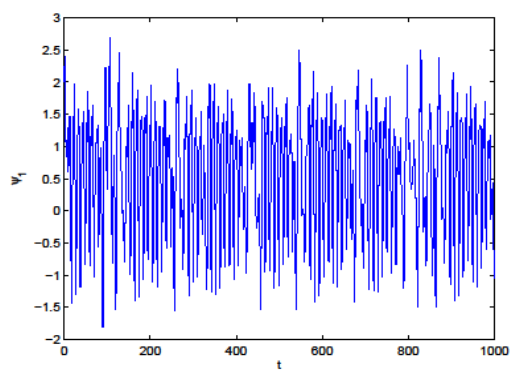

**Fig. 2** Dynamic of variable $\Psi_1$ for the set of parameters:

$\beta = 0.25, \alpha = 0.45, \tau = 0.35, i_{dc} = 0.3, i_0 = 0.7, \Omega = 0.6, \theta = 0$. Dimensionless units.

Fig. 3 demonstrates error dynamics $\Psi_1 - \Psi_2$. After short transients synchronization error approaches zero.

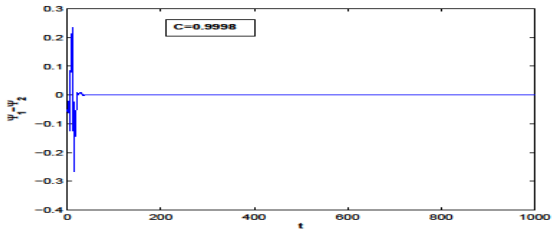

**Fig.3** Difference $\Psi_1 - \Psi_2$ versus t for the set of parameters as in **Fig.2**. Correlation coefficient between variables $\Psi_1$ and $\Psi_2$ C=0.9998. Dimensionless units.

Fig.4 shows dependence between variables $\Psi_2$ and $\Psi_3$.

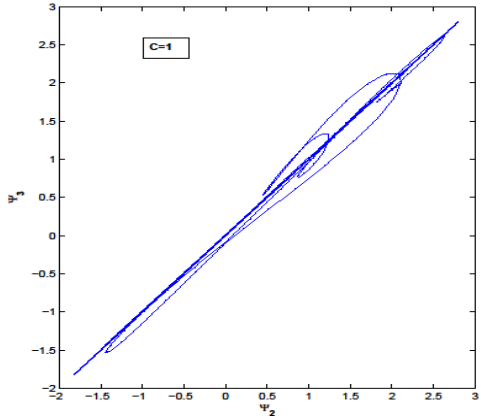

**Fig. 4** Dependence between variables $\Psi_2$ and $\Psi_3$. This dependence is linear: $\Psi_2 = \Psi_3$ in accordance with the complete synchronization between Terahertz Lasers 2 and 3. Correlation coefficient between $\Psi_2$ and $\Psi_3$ C=1. Parameters are as in **Fig.2** Dimensionless units.

Chaotic dynamics is of paramount importance in chaos based communication. Message can be hidden in chaotic envelope. As synchronization means communication chaos synchronization is used in reliable message exchange. Due to the presence of chaos data exchange security can be enhanced. Additionally, synchronization is of vital importance in achieving high power laser, including Terahertz lasers. This could give rise to the power level (around 1 milli Watt) needed for the application purposes.

## 4 Conclusions

In this work we have numerically shown the principal possibility of near perfect synchronization between the Terahertz lasers. We have spanned the bridge between chaos synchronization and widely used network topology- the ring topology. As mentioned above synchronization is vital in chaos-based communication system to decode the transmitted message [1]. Chaos based communication approach could provide an extra layer of security in data exchange between communicating computers and between Terahertz lasers, etc.

There are many types of network topologies: ring, star, mesh, hybrid, etc. Networks are also different from the viewpoint of applications, such as local area networks, wide area networks, etc. Security is also an issue for the different types of networks along with cost effectiveness. Data transfer speed is of immense important from the application point of view.

Here we dwell on the advantages and disadvantages of the ring topology in the computer network(s) architecture [22]. Studied in this paper the ring topology is used in the local area networks, wide area networks and synchronous optical fiber networks. Among advantages of the ring topology one can emphasize [22]:1) as data flows in one direction, this configuration reduces the chance of data packets' collision; 2) high speed communication up to 100 Mb/s; synchronous optical fiber networks, which could provide communication speed around 40Gb/s. 3) cost effectiveness and easy to manage, etc. It is also worth noting the following drawbacks [22]: troubleshooting is difficult; 2) increased number of nodes delays communications; 3) security is not high, etc.

Terahertz communications [24] means: effective data rates exchange exceeding 1Tbit/s (usually on an optical carrier); communication with a Terahertz carrier wave [24]. In the case of optical cables between the networks data rates could be even higher than 1Tbit/s. This is due to the fact that wave division multiplexing approach [25] can be exploited. In this approach data packets can be sent using multiple wavelengths.

# References

1. Schoell,E., Schuster,H.G.: Editors (second edition) Handbook of Chaos Control. (Wiley-VCH, Weinheim, 2008). https://doi.org/10.1002/9783527622313

2. Shahverdiev,E.M.: Modulated time delays, synchronized Josephson junctions in high temperature superconductors and chaotic terahertz waves. Journal of Superconductivity and Novel Magnetism 34, 1125 -1132 (2021). https://doi.org/10.1007/s10948-021-05837-7

3. Shahverdiev,E.M., Bayramov, P.A., Nuriev, R.A., Qocayeva,M.V., Hashimova, L.H.: Chaos synchronization between Josephson junctions coupled in series and driven by a central junction. Physica C: Superconductivity and its applications 557, 26-32 (2019). https://doi.org/10.1016/j.physc.2019.03

4. Shahverdiev,E.M.: Effect of parameter mismatches on chaos synchronization between Josephson junctions coupled in series and driven by a central junction. Physica C:Superconductivity and its applications 561, 52-57 (2019). https://doi.org/10.1016/j.physc.2019.03

5. Nishijima, S., Eckroad ,S, Marian, A., Choi, A.K., Kim, W.S., Terai, M.,Deng,Z., Zheng, J., Wang, J., Umemoto, K., Du, J., Febvre, P. , Keenan, S., Mukhanov, O., Cooley, L.D., Foley, C.P., Hassenzahl, V., Izumi, M.: Superconductivity and the environment: A Roadmap. Superconductor Science and Technology 26, 22-24 (2013). https://doi.org/10.1088/0953-2048/26/11/113001

6. Pecora, L.M., Carroll, T.L. :Synchronization in chaotic systems. Phys Rev Lett 64, 821-824 (1990). https://doi.org/10.1103/PhysRevLett.64.821

7. Shahverdiev, E.M. :Boundedness of dynamical systems and chaos synchronization. Phys. Rev. E 60, 3905-3909 (1999). https://doi.org/10.1103/PhysRevE.60.3905

8. Rosenblum,M.,Pikovsky, A.,Kurths,J.: From phase to lag synchronization in coupled chaotic oscillators. Phys. Rev. Lett.78, 4193-4196 (1997). https://doi.org/10.1103/PhysRevLett.78.4193.

9. Shahverdiev, E.M., Sivaprakasam,S., Shore, K.A.: Lag synchronization in time–delayed systems. Phys.Lett.A 292, 320-324 (2002). https://doi.org/10.1016/S0375-9601(01)00824-6.

10. Shahverdiev, E.M., Nuriev, R.A., Hashimova,R.H., Huseynova E.M.

Hashimov, R.H. , Shore, K.A.: Complete inverse chaos synchronization, parameter mismatches and generalized synchronization in the multi-feedback Ikeda model.Chaos , Solitons and Fractals 36, 211-216 (2008). https://doi.org/10.1016/j.chaos.2006.06.026.

11. Ghosh,S., Sar, G.K., Majhi, S., Ghosh,D.: Antiphase synchronization in a population of swarmalators, Phys. Rev. E 108 , 034217 (2023). https://doi.org/10.1103/PhysRevE.108.034217

12. Shahverdiev,E.M., Shore, K.A.:Generalized synchronization in time delayed systems. Phys.Rev.E 71 (2005) 016201. https://doi.org/10.1103/PhysRevE.71.016201

13 Shahverdiev, E.M., Bayramov,P.A., Shore, K.A.:Cascaded and adaptive chaos synchronization in multiple time-delay laser systems. Chaos, Solitons and Fractals, 180-186 (2009). https://doi.org/10.1016/j.chaos.2008.11.004

14. Shahverdiev,E.M., Sivaprakasam, S., Shore K.A.:Dual and dual-cross synchronization in chaotic systems. Optics Communications 216, 179-183(2003). https://doi.org/10.1016/S0030-4018(02)02286-1

15. Voss, H.U.: Anticipating chaotic synchronization Phys. Rev. E 61, 5115-5119 (2005). https://doi.org/10.1103/PhysRevE.61.5115.

16. Sivaprakasam,S., Shahverdiev,E.M., Spencer,P.S., Shore, K.A. :Experimental demonstration of anticipating synchronization in chaotic semiconductor lasers with optical feedback. Phys. Rev. Lett. 87, 154101 (2001). https://doi.org/10.1103/PhysRevLett.87.154101

17. Ott, E., Spano,M. :Controlling chaos. Physics Today 48, 34-40 (1995). https://doi.org/10.1063/1.881461

18. Ditto,W.L.,Showalter,K.: Introduction: Control and synchronization of chaos.Chaos 7, 509-511 (1997). https://doi.org/10.1063/1.166276.

19. Yanchuk,S., G. Giacomelli, G. Spatio-temporal phenomena in complex systems with time delays. J. of Phys. A: Mathematical and Theoretical 30 103001(2017). https://doi.org/10.1088/1751-8121/50/10/103001.

20. Shahverdiev,E.M., Shore K.A.: Chaos synchronization regimes in multiple time delay semiconductor lasers. Phys.Rev.E 77, 057201 (2008). https://doi.org/10.1103/PhysRevE.77.057201.

21. Shahverdiev,E.M, :Synchronization in systems with multiple time delays. Phys.Rev.E 70, 067202 (2004).https://doi.org/10.1103/PhysRevE.70.067202

22. https://www.wikipedia.org/Ring_network, Accessed 25 May 2025.

23. Strogatz, S.H. :Nonlinear Dynamics and Chaos:with applications to Physics, Biology, Chemistry, and Engineering. (Addison-Wesley, Reading ,2014). https://doi.org/10.1201/9780429492563

24. Fitch, M.J., Osiander, R.:Terahertz waves for communications and sensing. Johns Hopkins APL Technical digest 25**,** 348-355 (2004)

25. Argyris,A.,Syvridis,D., Larger,L., Annovazi-Lodi, A., Colet, P., Fischer,I,, Garsio-Ojalvo,J., Mirasso,C.R., Pesquera,L. Shore, K.A. :Chaos- based communications at high bit rates using commercial fibre-optic links. Nature 438, 343-346 (2005). https://doi.org/10.1038/nature04275

.